\documentclass[%
 reprint,
superscriptaddress,
showpacs,preprintnumbers,
 amsmath,amssymb,
 aps,
 prl,
]{revtex4-1}
\usepackage{graphicx}
\usepackage{dcolumn}
\usepackage{bm}

\begin{document}

\title{Nuclear Magnetic Resonance for Arbitrary Spin Values in the Rotating Wave Approximation}

\author{Zhichen Liu}
\affiliation{Department of Physics, University of Central Florida, Orlando, FL 32816-2385, USA}
\author{Sunghyun Kim}
\affiliation{Department of Physics, University of Central Florida, Orlando, FL 32816-2385, USA}
\author{Richard A. Klemm}
\email{richard.klemm@ucf.edu, corresponding author}
\affiliation{Department of Physics, University of Central Florida, Orlando, FL 32816-2385, USA}
\affiliation{U. S. Air Force Research Laboratory, Wright-Patterson Air Force Base, Ohio 45433-7251, USA}
\date{\today}

\begin{abstract}
In order to probe the transitions of a nuclear spin $s$ from one of it substate quantum numbers $m$ to another substate $m'$, the experimenter applies a magnetic field ${\bm B}_0$ in some particular direction, such along $\hat{\bm z}$, and then applies an weaker field ${\bm B}_1(t)$ that is oscillatory in time with the angular frequency $\omega$, and is normally perpendicular to ${\bm B}_0$, such as ${\bm B}_1(t)=B_1\hat{\bm x}\cos(\omega t)$.  In the rotating wave approximation, ${\bm B}_1(t)=B_1[\hat{\bm x}\cos(\omega t)+\hat{\bm y}\sin(\omega t)]$.  Although this problem is solved for spin $\frac{1}{2}$ in every quantum mechanics textbook, for the general spin $s$ case, its general solution has been published only for the overall probability of a transition between the states, but the time dependence of the probability of finding the nucleus in each of the substates has not previously been published.  Here we present an elementary method to solve this problem exactly, and present figures for the time dependencies of the various substates states for a variety of initial substate probabilities for a variety of $s$ values. We found a new result:  unlike the $s=\frac{1}{2}$ case, for which if the initial probability of finding the particle in one of the substates was 1, and the time dependence of the probabilities of each of the substates oscillates between 0 and 1, for higher spin values, the time dependencies of the probabilities finding the particle in each of its substates, which periodic, is considerably more complicated.
\end{abstract}

 \vskip0pt
\maketitle

\section{Introduction}
Nuclear dipole moments have spin quantum numbers $s$ that vary from 0 to 9/2 for stable nuclei isotopes of Nb, Sr, and Hf, and can be much larger than that for long-lived nuclei, such as $^{166}_{67}$Ho, with a lifetime of 1200 years and a spin of 7, and for $^{213}_{87}$Fr, with a lifetime of 3.1 $\mu$s, with spin 65/2 \cite{Stone}.  Therefore there has long been great interest in magnetic resonance experiments on nuclei of arbitrary spin values.
In a standard magnetic resonance experiment \cite{Majorana,RabiRamseySchwinger,Ramsey} the applied magnetic field ${\bm B}(t)$ has a form similar to
\begin{eqnarray}
{\bm B}(t)&=&B_0\hat{\bm z}+B_1\hat{\bm x}\cos(\omega t).
\end{eqnarray}
although the exact solution to this problem has never been published, even for spin $\frac{1}{2}$, those authors found that one could make great progress by use of the rotating wave approximation, in which the oscillatory field component rotates about the constant field component,
\begin{eqnarray}
{\bm B}(t)&=&B_0\hat{\bm z}+B_1[\hat{\bm x}\cos(\omega t)+\hat{\bm y}\sin(\omega t)],
\end{eqnarray}
where usually one has $B_1/B_0\ll 1$.

That is, one can treat the standard model as the sum of two models
\begin{eqnarray}
{\bm B}_{\pm}(t)&=&B_0\hat{\bm z}+\frac{B_1}{2}[\hat{\bm x}\cos(\omega t)\pm\hat{\bm y}\sin(\omega t)],
\end{eqnarray}
which are rotating in the clockwise and anticlockwise senses, and then either add the results together, or treat one of them as a perturbation \cite{Majorana,RabiRamseySchwinger,Ramsey,Gottfried}.  Although the first treatment of the problem in the rotating wave approximation by Majorana \cite{Majorana} was  correct, the probability of a transition was calculated, but the fully time dependence of the wave function was not obtained.  Similar results were obtained by Rabi, Ramsey, and Schwinger \cite{RabiRamseySchwinger}, and in Appendix E of Ramsey's book \cite{Ramsey}, which the author attributed to Schwinger, the $|s=I,m\rangle$ state was constructed from $2I$ Pauli matrices. Although such as method was shown to give the correct answer for half-integral spins, and correctly gave the correct probability of a transition from one substate to another, that method did not apply to integral spin values. In both treatments by Majorana and Schwinger in Ramsey's book \cite{Majorana,Ramsey}, the derivation for arbitrary spin value was complicated, and the time dependence of the occupation probability for each of the spin states $|s,m\rangle$ was not given.

The Hamiltonian for these constant and  oscillatory magnetic fields in the rotating wave approximation is therefore
\begin{eqnarray}
H&=&\omega_0S_z+\omega_1[S_x\cos(\omega t)+S_y\sin(\omega t)],
\end{eqnarray}
where
\begin{eqnarray}
\omega_0&=&\mu_NB_0,\nonumber\\
\omega_1&=&\mu_NB_1,
\end{eqnarray}
where $\mu_N$ is the nuclear magneton appropriate for the nucleus under study.

 In Gottfried's book \cite{Gottfried}, he presented a much simpler way to solve the problem in the rotating wave approximation, by performing a rotation about the $z$ axis by the {\it time-dependent}  ``angle'' $\omega t$.  However, he then wrote the wave function for the general state in the form
\begin{eqnarray}
|\psi(t)\rangle&=&e^{-iS_z\omega t/\hbar}e^{-iH_{\rm eff}t/\hbar}|\psi(0)\rangle,\label{psi1oft}
\end{eqnarray}
where
\begin{eqnarray}
H_{\rm eff}&=&(\omega_0-\omega)S_z+\omega_1S_x.
\end{eqnarray}
Although in Eq.(\ref{psi1oft}), the operator on the left containng $S_z$ is diagonal in the spin states $|s,m\rangle$, the second operator in $H_{\rm eff}$  contains both $S_z$ and $S_x$, and is therefore not diagonal, and he did not attempt to diagonalize it.  Here we use the same technique as those above-mentioned pioneers in the field \cite{Majorana,RabiRamseySchwinger,Ramsey,Gottfried}, but diagonalize the second term exactly.  The technique is amazingly simple, and allows for an exact expression for the time dependence of the amplitude for each of the substates of the wave function, as detailed in the following.

We could also write $\omega_0\rightarrow\omega_1\cos\alpha$ and $\omega_1\rightarrow\omega_1\sin\alpha$ as in Griffiths and Schroeter \cite{Griffiths}, which they described as the ``sweeping field'' model.
Both Griffiths and Schroeter and Sakurai and Napolitano \cite{Griffiths,SN3} only discussed the simplest case for spin $\frac{1}{2}$, and the former derivation was much more complicated than that done here for general spin values.

We note that the components of the spin operators satisfy the Lie algebra
\begin{eqnarray}
[S_i,S_j]&=&i\hbar\epsilon_{ijk}S_k,
\end{eqnarray}
where $\epsilon_{ijk}$ is the Levi-Civita symbol, repeated indices are summed over, and where $\epsilon_{xyz}=+1$ plus cyclic permutations, and $\epsilon_{yxz}=-1$ plus cyclic permutations.  We
first investigate a rotation of $S_x$ and $S_y$ by the angle $\phi$ about the $z$ axis.  We have
\begin{eqnarray}
S_x(\phi)&=&e^{iS_z\phi/\hbar}S_xe^{-iS_z\phi/\hbar},\nonumber\\
S_y(\phi)&=&e^{iS_z\phi/\hbar}S_ye^{-iS_z\phi/\hbar}.
\end{eqnarray}
We then obtain
\begin{eqnarray}
\frac{\partial S_x}{\partial \phi}&=&\frac{i}{\hbar}e^{iS_z\phi/\hbar}[S_z,S_x]e^{-iS_z\phi/\hbar}\nonumber\\
&=&-S_y(\phi),\\
\frac{\partial S_y}{\partial \phi}&=&\frac{i}{\hbar}e^{iS_z\phi/\hbar}[S_z,S_y]e^{-iS_z\phi/\hbar}\nonumber\\
&=&S_x(\phi),
\end{eqnarray}
Differentiating both of these equations with respect to $\phi$, we have
\begin{eqnarray}
\frac{\partial^2S_x(\phi)}{\partial\phi^2}+S_x(\phi)&=&0,\nonumber\\
\frac{\partial^2S_y(\phi)}{\partial\phi^2}+S_y(\phi)&=&0,
\end{eqnarray}
both of which are the classical one-dimensional harmonic oscillator equations, with solutions
\begin{eqnarray}
S_x(\phi)&=&A\cos(\phi)+B\sin(\phi),\label{Sx}\\
S_y(\phi)&=&C\cos(\phi)+D\sin(\phi),\label{Sy}
\end{eqnarray}
where $A,B,C,D$ are operators and $\phi$ is a real dimensionless constant.  Setting $S_x(0)=S_x$, $S_y(0)=S_y$, and from Eqs. (\ref{Sx}) and (\ref{Sy}),
\begin{eqnarray}
\frac{\partial S_x(\phi)}{\partial\phi}\Big|_{\phi=0}&=&-S_y,\nonumber\\
\frac{\partial S_y(\phi)}{\partial\phi}\Big|_{\phi=0}&=&S_x,
\end{eqnarray}
we easily find that
\begin{eqnarray}
S_x(\phi)&=&S_x\cos(\phi)-S_y\sin(\phi),\nonumber\\
S_y(\phi)&=&S_y\cos(\phi)+S_x\sin(\phi),
\end{eqnarray}
which behaves exactly as in a classical rotation of the $x$ and $y$ axes by the angle $\phi$ about the $z$ axis.
Letting $\phi\rightarrow -\omega t$, we have
\begin{eqnarray}
S_x(-\omega t)&=&S_x\cos(\omega t)+S_y\sin(\omega t)\nonumber\\
&=&e^{-iS_z\omega t/\hbar}S_xe^{iS_z\omega t/\hbar}.
\end{eqnarray}
The Hamiltonian may then be written as
\begin{eqnarray}
H&=&\omega_0S_z+\Bigl(e^{-iS_z\omega t/\hbar}\omega_1S_xe^{iS_z\omega t/\hbar}\Bigr)\nonumber\\
&=&e^{-iS_z\omega t/\hbar}\Bigl(\omega_0S_z+\omega_1S_x\Bigr)e^{iS_z\omega t/\hbar},
\end{eqnarray}
since $[S_z,S_z]=0$.
We then let
\begin{eqnarray}
|\psi\rangle&=&e^{-iS_z\omega t/\hbar}|\psi'\rangle,\nonumber\\
i\hbar\frac{\partial|\psi\rangle}{\partial t}&=&\omega S_ze^{-iS_z\omega t/\hbar}|\psi'\rangle+i\hbar e^{-iS_z\omega t/\hbar}\frac{\partial|\psi'\rangle}{\partial t},
\end{eqnarray}
From the Schr{\"o}dinger equation for the quantum spin wave function $|\psi\rangle$,
\begin{eqnarray}
H|\psi\rangle&=&i\hbar\frac{\partial|\psi\rangle}{\partial t},
\end{eqnarray}
we have
\begin{eqnarray}
H|\psi\rangle&=&He^{-iS_z\omega t/\hbar}|\psi'\rangle\nonumber\\
e^{-iS_z\omega t/\hbar}\Bigl(\omega_0S_z+\omega_1S_x\Bigr)|\psi'\rangle&=&e^{-iS_z\omega t/\hbar}\Bigl(\omega S_z|\psi'\rangle\nonumber\\
& &+i\hbar\frac{\partial|\psi'\rangle}{\partial t}\Bigr),\nonumber\\
e^{-iS_z\omega t/\hbar}\Bigl[(\omega_0-\omega)S_z+\omega_1S_x\Bigr]|\psi'\rangle&=&i\hbar e^{-iS_z\omega t/\hbar}\frac{\partial|\psi'\rangle}{\partial t}.\nonumber\\
\end{eqnarray}
Multiplying both sides of this equation by $\exp[iS_z\omega t/\hbar]$ on the left, we have
\begin{eqnarray}
H_{\rm eff}|\psi'\rangle&=&[(\omega_0-\omega)S_z+\omega_1S_x]|\psi'\rangle=i\hbar\frac{\partial|\psi'\rangle}{\partial t},
\end{eqnarray}
where the effective Hamiltonian $H_{\rm eff}$ is independent of the time $t$.
We next solve the time dependence of the Schr{\"o}dinger equation for $|\psi'\rangle$, obtaining
\begin{eqnarray}
|\psi'(t)\rangle&=&e^{-iH_{\rm eff}t/\hbar}|\psi'(0)\rangle,
\end{eqnarray}
and this leads to Eq. (\ref{psi1oft}),
precisely as found by Gottfried \cite{Gottfried}.  However, as mentioned in the introduction,  Gottfried didn't finish the problem, as he didn't diagonalize $H_{\rm eff}$.

Since the spin ${\bm S}$ lies in the $xz$ plane, we let $\hat{\bm u}$ be a unit vector parallel to ${\bm S}$,
\begin{eqnarray}
\hat{\bm u}_z&=&\hat{\bm z}\frac{(\omega_0-\omega)}{\Omega},\\
\hat{\bm u}_x&=&\hat{\bm x}\frac{\omega_1}{\Omega},
\end{eqnarray}
and normalization of the unit vector requires
\begin{eqnarray}
\hat{\bm u}\cdot\hat{\bm u}&=&1,\nonumber\\
\frac{(\omega_0-\omega)^2}{\Omega^2}+\frac{\omega_1^2}{\Omega^2}&=&1,\nonumber\\
\Omega&=&\sqrt{(\omega_0-\omega)^2+\omega_1^2}.\label{Omega}
\end{eqnarray}
We then perform a rotation of  ${\bm S}$ about the $y$ axis by the angle $\beta$, so that after the rotation, it will point in the $z$ axis direction.  We thus let
\begin{eqnarray}
|\psi(t)\rangle&=&e^{-iS_z\omega t/\hbar}e^{-i\beta S_y/\hbar}e^{+i\beta S_y/\hbar}e^{-iH_{\rm eff}t/\hbar}\nonumber\\
& &\times e^{-i\beta S_y/\hbar}e^{+i\beta S_y/\hbar}|\psi(0)\rangle\\
&=&e^{-iS_z\omega t/\hbar}e^{-i\beta S_y/\hbar}e^{-i\tilde{H}_{\rm eff}(\beta)t/\hbar}e^{i\beta S_y/\hbar}|\psi(0)\rangle,\label{psioft}\nonumber\\
\end{eqnarray}
where
\begin{eqnarray}
\tilde{H}_{\rm eff}(\beta)&=&e^{+i\beta S_y/\hbar}H_{\rm eff}e^{-i\beta S_y/\hbar},
\end{eqnarray}
and letting
\begin{eqnarray}
{\cal O}(\beta,t)&=&e^{+i\beta S_y/\hbar}e^{-iH_{\rm eff}t/\hbar}e^{-i\beta S_y/\hbar},
\end{eqnarray}
we have
\begin{eqnarray}
{\cal O}(\beta,t)&=&
e^{+i\beta S_y/\hbar}\sum_{n=0}^{\infty}\frac{(-iH_{\rm eff}t/\hbar)^n}{n!}e^{-i\beta S_y/\hbar}\nonumber\\
&=&1+\frac{1}{1!}e^{+i\beta S_y/\hbar}(-iH_{\rm eff}t/\hbar)e^{-i\beta S_y/\hbar}\nonumber\\
& &+\frac{1}2!e^{+i\beta S_y/\hbar}(-iH_{\rm eff}t/\hbar)e^{-i\beta S_y/\hbar}\nonumber\\
& &\times e^{+i\beta S_y/\hbar}(-iH_{\rm eff}t/\hbar)e^{-i\beta S_y/\hbar}+\dots\nonumber\\
&=&1+\frac{1}{1!}(-i\tilde{H}_{\rm eff}(\beta)t/\hbar)+\frac{1}{2!}[-i\tilde{H}_{\rm eff}(\beta)t/\hbar]^2+\ldots\nonumber\\
&=&e^{-i\tilde{H}_{\rm eff}(\beta)t/\hbar}.
\end{eqnarray}
We then have
\begin{eqnarray}
S_z(\beta)&=&e^{iS_y\beta/\hbar}S_ze^{-iS_y\beta/\hbar},\nonumber\\
S_x(\beta)&=&e^{iS_y\beta/\hbar}S_xe^{-iS_y\beta/\hbar}.
\end{eqnarray}
As for the rotation of $S_x$ and $S_y$ by $\phi$ about the $z$ axis, we find for rotations by the angle $\beta$ about the $y$ axis that
\begin{eqnarray}
S_x(\beta)&=&S_x\cos(\beta)+S_z\sin(\beta),\\
S_z(\beta)&=&S_z\cos(\beta)-S_x\sin(\beta),
\end{eqnarray}
and
\begin{eqnarray}
\tilde{H}_{\rm eff}(\beta)&=&(\omega_0-\omega)S_z(\beta)+\omega_1S_x(\beta)\nonumber\\
&=&(\omega_0-\omega)[S_z\cos(\beta)-S_x\sin(\beta)]\nonumber\\
& &+\omega_1[S_x\cos(\beta)+S_z\sin(\beta)].
\end{eqnarray}
Forcing the coefficient of $S_x$ to vanish, we have
\begin{eqnarray}
-(\omega_0-\omega)\sin(\beta)+\omega_1\cos(\beta)&=&0,\nonumber\\
\tan\beta&=&\frac{\omega_1}{\omega_0-\omega},
\end{eqnarray}
which implies
\begin{eqnarray}
\sin(\beta)&=&\frac{\omega_1}{\Omega},\label{sinbeta}\\
\cos(\beta)&=&\frac{\omega_0-\omega}{\Omega}.\label{cosbeta}
\end{eqnarray}
The rotated $\tilde{H}_{\rm eff}(\beta)$ then becomes
\begin{eqnarray}
\tilde{H}_{\rm eff}(\beta)&=&(\omega_0-\omega)S_z\cos(\beta)+\omega_1S_z\sin(\beta)\nonumber\\
&=&\frac{(\omega_0-\omega)^2}{\Omega}S_z+\frac{\omega_1^2}{\Omega}S_z\nonumber\\
&=&\Omega S_z,
\end{eqnarray}
which is diagonal in the $|s,m\rangle$ representation.

We then have from Eq. (\ref{psioft}) that
\begin{eqnarray}
|\psi(t)\rangle&=&e^{-iS_z\omega t/\hbar}e^{-i\beta S_y/\hbar}e^{-iS_z\Omega t/\hbar}e^{i\beta S_y/\hbar}|\psi(0)\rangle,\nonumber\\
\end{eqnarray}
where $\beta$ is the rotation angle determined by Eqs. (\ref{sinbeta}) and (\ref{cosbeta}).  Note that the overall operator consists of four exponential operators, none of which commutes with its neighbor or neighbors. However, this exponential form allows us to find an exact form for $|\psi(t)\rangle$ by introducing a complete set of outer products of the states between each pair of exponential operators.

We then let
\begin{eqnarray}
|\psi(0)\rangle&=&\sum_{m'=-s}^sC_{m'}(0)|s,m'\rangle,\\
|\psi(t)\rangle&=&\sum_{m''=-s}^sC_{m''}(t)|s,m''\rangle,
\end{eqnarray}
Now, we take the inner product of this last equation with the bra $\langle s,m|$, obtaining
\begin{eqnarray}
C_{m}(t)&=&\sum_{m'=-s}^sC_{m'}(0)\langle s,m|e^{-iS_z\omega t/\hbar}e^{iS_y\beta/\hbar}e^{-iS_z\Omega t/\hbar}\nonumber\\
& &\hskip60pt\times e^{-iS_y\beta/\hbar}|s,m'\rangle,
\end{eqnarray}
where we used the orthonormality of the states on the left-hand side,
\begin{eqnarray}
\langle s,m|s,m''\rangle&=&\delta_{m,m''}.
\end{eqnarray}
Inserting an identity operation such as
\begin{eqnarray}
\sum_{m'''=-s}^s|s,m'''\rangle\langle s,m'''|&=&1,
\end{eqnarray}
between each of the exponential operators,
we have
\begin{eqnarray}
C_m(t)&=&\sum_{m',m'',m''',m''''=-s}^sC_{m'}(0)\langle s,m|e^{-iS_z\omega t/\hbar}|s,m'''\rangle\nonumber\\
& &\times\langle s,m'''|e^{-i\beta S_y/\hbar}|s,m''\rangle\nonumber\\
& &\times\langle s,m''|e^{-iS_z\Omega t/\hbar}|s,m''''\rangle\langle s,m''''|e^{i\beta S_y/\hbar}|s,m'\rangle\nonumber\\
&=&\sum_{m',m''m''',m''''=-s}^sC_{m'}(0)e^{-im\omega t}\delta_{m,m'''}e^{-im''\Omega t}\nonumber\\
& &\times\delta_{m'',m''''}\langle s,m'''|e^{-i\beta S_y/\hbar}|s,m''\rangle\nonumber\\
 & &\times\langle s,m''''|e^{i\beta S_y/\hbar}|s,m'\rangle\nonumber\\
&=&e^{-im\omega t}\sum_{m',m''=-s}^sC_{m'}(0)e^{-im''\Omega t}\nonumber\\
& &\times\langle s,m|e^{-i\beta S_y/\hbar}|s,m''\rangle\langle s,m''|e^{i\beta S_y/\hbar}|s,m'\rangle\nonumber\\
&=&e^{-im\omega t}\sum_{m',m''=-s}^sC_{m'}(0)e^{-im''\Omega t}\nonumber\\
& &\hskip70pt\times d_{m,m''}^{(s)}(\beta)d_{m',m''}^{(s)*}(\beta),
\end{eqnarray}
where we made use of the diagonal operator equation
\begin{eqnarray}
S_z|s,m\rangle&=&\hbar m|s,m\rangle.
\end{eqnarray}
Of course we have
\begin{eqnarray}
\sum_{m=-s}^s|C_m(0)|^2&=&\sum_{m=-s}^s|C_m(t)|^2=1,\\
|C_m(t)|^2&=&\Biggl|\sum_{m',m''=-s}^sC_{m'}(0)e^{-im''\Omega t}\nonumber\\
& &\hskip30pt\times d^{(s)}_{m,m''}(\beta)d^{(s)*}_{m',m''}(\beta)\Biggr|^2. \label{probability}
\end{eqnarray}

In this formula, $d^{(s)}_{m',m}(\beta)$
is the reduced rotation matrix element for the second Euler angle, as derived by Wigner and described in detail in Sakurai and Napolitano \cite{SN3} and in less transparent form in Gottfried.  According to \cite{SN3}, it is
\begin{eqnarray}
d^{(s)}_{m',m}(\beta)&=&\sum_k(-1)^{k-m+m'}\nonumber\\
& &\times\frac{\sqrt{(s+m)!(s-m)!(s+m')!(s-m')!}}{(s+m-k)!k!(s-k-m')!(k-m+m')!}\nonumber\\
& &\times\Bigl[\cos(\beta/2)\Bigr]^{2s-2k+m-m'}\Bigl[\sin(\beta/2)\Bigr]^{2k-m+m'},\label{dsm'm}
\end{eqnarray}
where $k$ takes on all values for which none of the arguments of the factorials in the denominator are negative.  That is, ${\rm max}(0, m-m')\le k\le {\rm min}(s+m,s-m')$.  In the present case, we have
\begin{eqnarray}
\sin(\beta/2)&=&\sqrt{\frac{1-\cos(\beta)}{2}}=\sqrt{\frac{\Omega+\omega-\omega_0}{2\Omega}},\\
\cos(\beta/2)&=&\sqrt{\frac{1+\cos(\beta)}{2}}=\sqrt{\frac{\Omega+\omega_0-\omega}{2\Omega}},
\end{eqnarray}
which can be easily inserted into the final form, Eq. (\ref{dsm'm}.

We note that in the adiabatic approximation, as discussed elsewhere \cite{GY}, there are no transitions from one state to another, so that
\begin{eqnarray}
C_m^{\rm ad}(t)&=&C_m(0)e^{-im\omega t}e^{-im\Omega t}|d^{(s)}_{m,m}(\beta)|^2\label{adiabatic}
\end{eqnarray}
so that the dynamic phase is the first exponential, and the Berry phase $m\Omega T$ is obtained from the second exponential at the time $T$ for the completion of a full period $T$ of the motion \cite{Griffiths,SN3,GY}.

We also note that the time dependence of this has precisely the same form as that of Hall and Klemm\cite{HallKlemm}
\begin{eqnarray}
|I,m_I\rangle(t)&=&e^{im_I\omega_0t}\sum_{m_I'=-I}^{I}C_{m_I'}^{m_I}e^{im_I'\Gamma_nt},
\end{eqnarray}
where $s=I$, $m_I=m$, $m_I'=m'$, $\Gamma_n=\Omega$, $\omega_0=\omega$, $\omega_n=\omega_0$, and $\Omega_n=\omega_1$.
 Although they only showed it to be true for $s=1/2,1,3/2$, and did not provide specific formulas for the $C_{m_I'}^{m_I}$, it is now shown to be true for general nuclear spin $I$, and the
 \begin{eqnarray}
 C_{m_I'}^{m_I}&=&C_{m_{I'}}(0)\sum_{m_{I''}=-I}^Id^{(I)}_{m_{I},m_{I''}}(\beta)d_{m_{I'},m_{I''}}^{(I)*}(\beta)
  \end{eqnarray}
  in the present notation.

  In Figs. 1 and 2, we present  plots of $|C_m(\tau)|$ of the spin 1/2 states with three different initial conditions, where  $\tau=t/T$ is the reduced time, and the period of the motion $T=2\pi/\Omega$, where $\Omega$ is given by Eq. (\ref{Omega}). We generally take $\omega_1/\omega_0=0.01$, which could be experimentally relevant.  In each figure, one period of the motion is presented.    In these figures, we display the results on resonance, $\omega=\omega_0$, in the resonance peak at $\omega=\omega_0+\omega_1$, and $\omega=\omega_0+3\omega_1$, which is well off resonance.
  \begin{figure}
  \center{\includegraphics[width=0.49\textwidth]{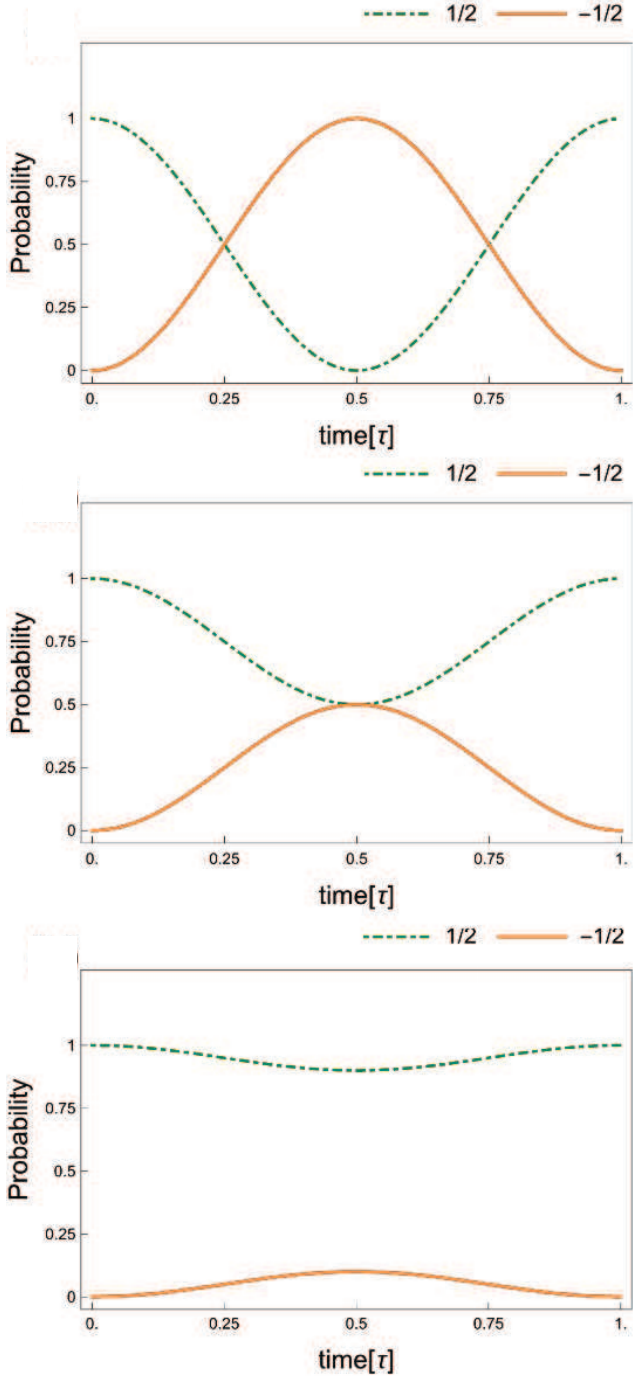}
  \caption{Plots for $s=\frac{1}{2}$ of the $|C_m(\tau)|^2$ from Eq. (\ref{probability}), where  the reduced time variable $\tau=t/T$, the period $T=2\pi/\Omega$, and $\Omega$ is given by Eq. (\ref{Omega}).  In each figure, $|C_{1/2}(\tau)|^2$ is dashed green, $|C_{-1/2}(\tau)|^2$ is solid orange, and $|C_{1/2}(0)|^2=1$, $|C_{-1/2}(0)|^2=0$.  Top:  $\omega=\omega_0$.  Middle:  $\omega=\omega_0+\omega_1$. Bottom: $\omega=\omega_0+3\omega_1$.}}
  \end{figure}

     \begin{figure}
  \center{\includegraphics[width=0.49\textwidth]{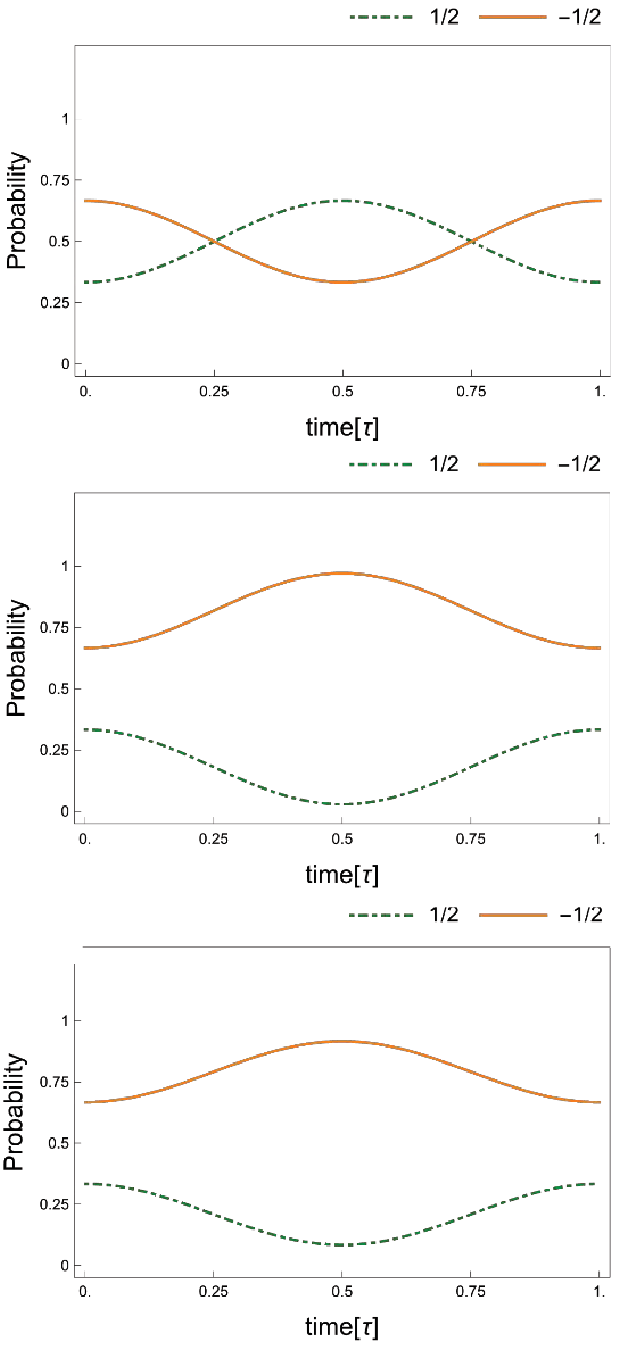}
  \caption{Plots for $s=\frac{1}{2}$ of the $|C_m(\tau)|^2$ from Eq. (\ref{probability}), where  the reduced time variable $\tau=t/T$, the period $T=2\pi/\Omega$, and $\Omega$ is given by Eq. (\ref{Omega}).  In each figure, $|C_{1/2}(\tau)|^2$ is dashed green, $|C_{-1/2}(\tau)|^2$ is solid orange, and $|C_{1/2}(0)|^2=\frac{1}{3}$, $|C_{-1/2}(0)|^2=\frac{2}{3}$.  Top:  $\omega=\omega_0$.  Middle:  $\omega=\omega_0+\omega_1$. Bottom: $\omega=\omega_0+3\omega_1$.}}
  \end{figure}

     \begin{figure}
  \center{\includegraphics[width=0.49\textwidth]{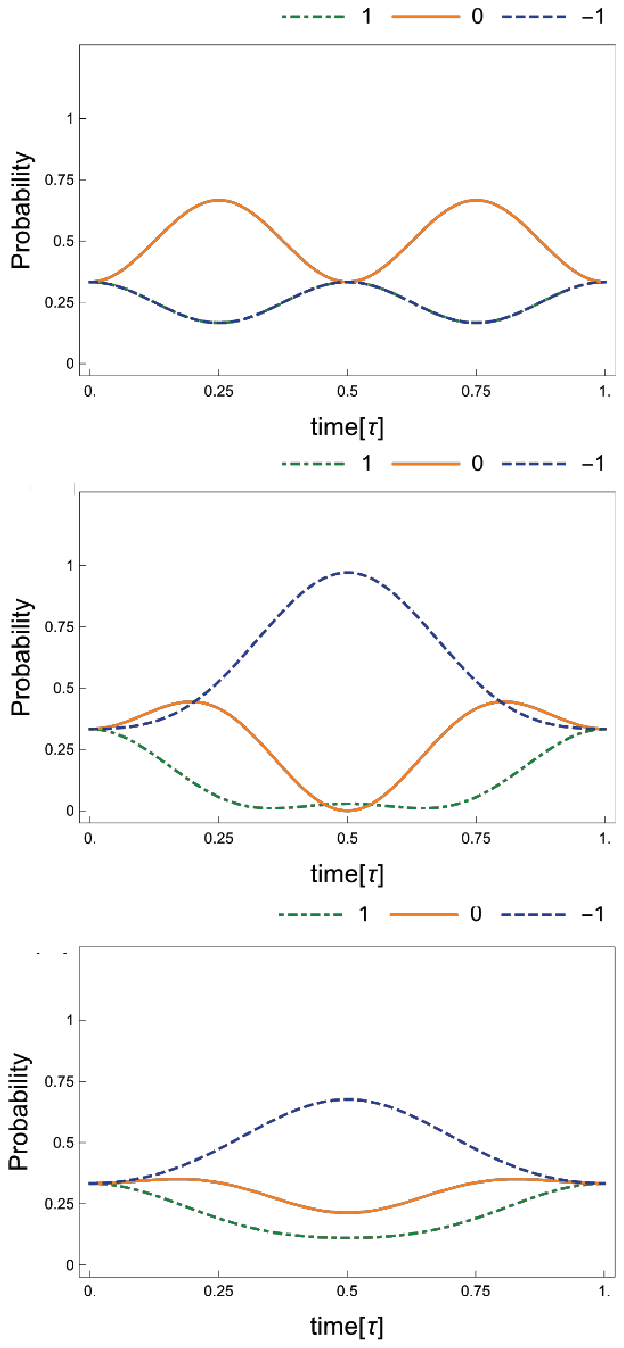}
  \caption{Plots for $s=1$ of the $|C_m(\tau)|^2$ from Eq. (\ref{probability}), where  the reduced time variable $\tau=t/T$, the period $T=2\pi/\Omega$, and $\Omega$ is given by Eq. (\ref{Omega}).  In each figure, $|C_{1}(\tau)|^2$ is dot dashed green, $|C_{0}(\tau)|^2$ is solid orange, and $C_{-1}(\tau)|^2$ is dashed blue, and $|C_{m}(0)|^2=\frac{1}{3}$.  Top:  $\omega=\omega_0$.  Middle:  $\omega=\omega_0+\omega_1$. Bottom: $\omega=\omega_0+3\omega_1$.}}
  \end{figure}
  
     \begin{figure}
  \center{\includegraphics[width=0.49\textwidth]{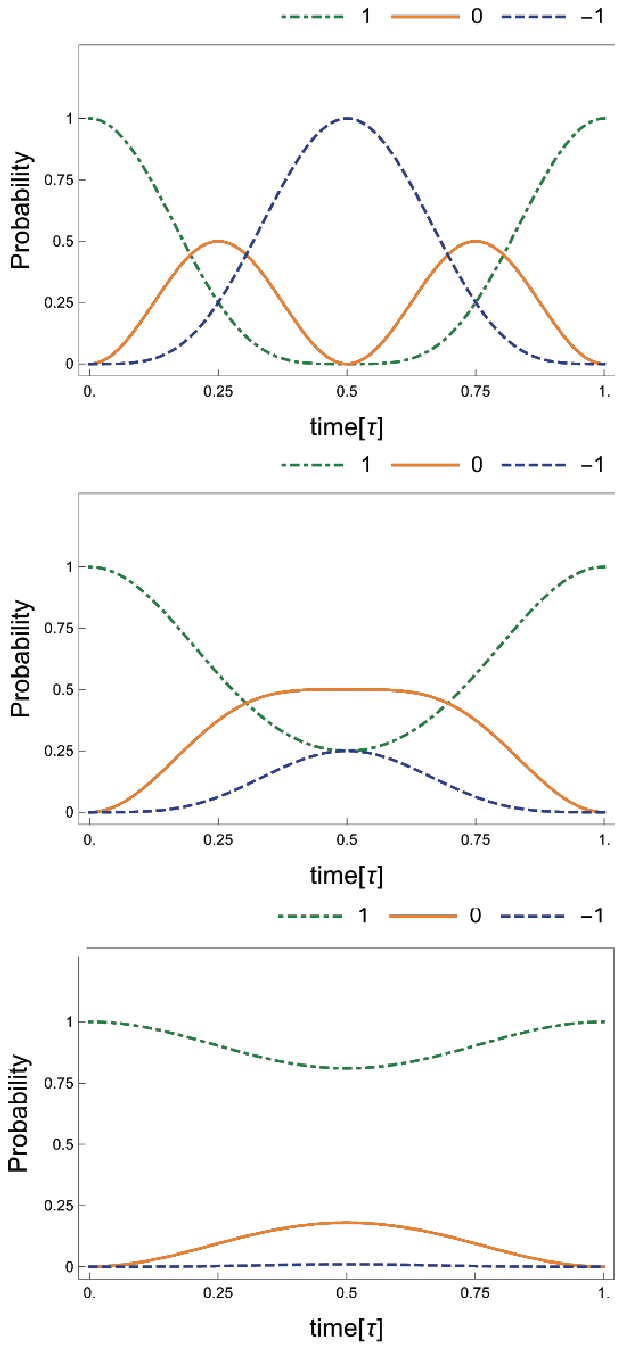}
  \caption{Plots for $s=1$ of the $|C_m(\tau)|^2$ from Eq. (\ref{probability}), where  the reduced time variable $\tau=t/T$, the period $T=2\pi/\Omega$, and $\Omega$ is given by Eq. (\ref{Omega}).  In each figure, $|C_{1}(\tau)|^2$ is dot dashed green, $|C_{0}(\tau)|^2$ is solid orange, and $|C_{-1}(\tau)|^2$ is dashed blue, and $|C_{1}(0)|^2=1$ and $|C_0(0)|^2=|C_{-1}(0)|^2=0$.  Top:  $\omega=\omega_0$.  Middle:  $\omega=\omega_0+\omega_1$. Bottom: $\omega=\omega_0+3\omega_1$.}}
  \end{figure}
      \begin{figure}
  \center{\includegraphics[width=0.49\textwidth]{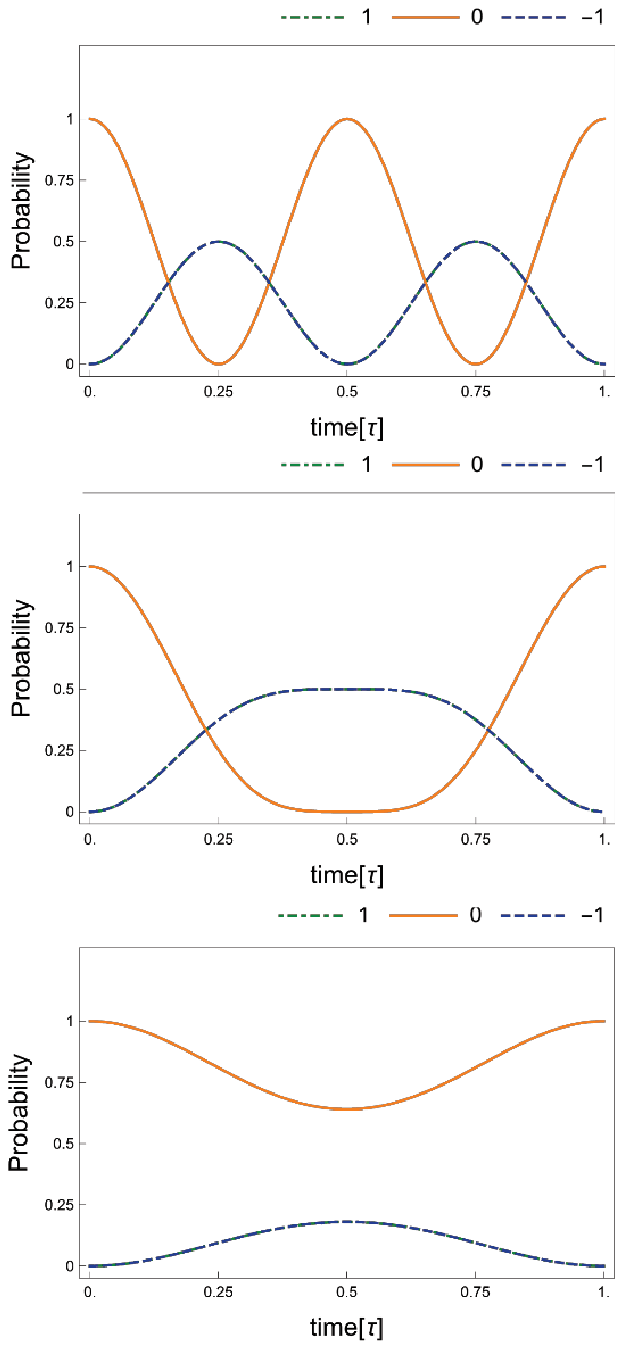}
  \caption{Plots for $s=1$ of the $|C_m(\tau)|^2$ from Eq. (\ref{probability}), where  the reduced time variable $\tau=t/T$, the period $T=2\pi/\Omega$, and $\Omega$ is given by Eq. (\ref{Omega}).  In each figure, $|C_{1}(\tau)|^2$ is dot dashed green, $|C_{0}(\tau)|^2$ is solid orange, and $|C_{-1}(\tau)|^2$ is dashed blue, and $|C_{0}(0)|^2=1$ and $|C_1(0)|^2=|C_{-1}(0)|^2=0$.  Top:  $\omega=\omega_0$.  Middle:  $\omega=\omega_0+\omega_1$. Bottom: $\omega=\omega_0+3\omega_1$.}}
  \end{figure}
      \begin{figure}
  \center{\includegraphics[width=0.49\textwidth]{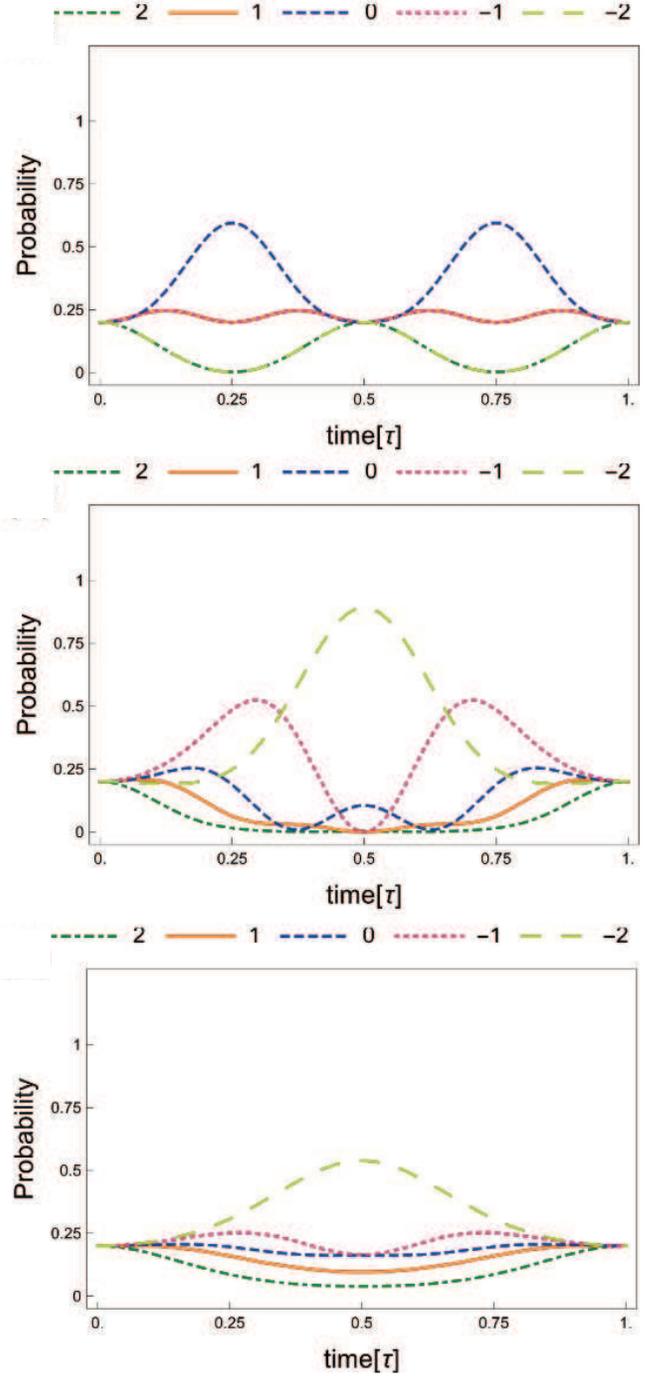}
  \caption{Plots for $s=2$ of the $|C_m(\tau)|^2$ from Eq. (\ref{probability}), where  the reduced time variable $\tau=t/T$, the period $T=2\pi/\Omega$, and $\Omega$ is given by Eq. (\ref{Omega}).  In each figure, $|C_{2}(\tau)|^2$ is dot dashed green, $|C_{1}(\tau)|^2$ is solid orange, and $|C_{0}(\tau)|^2$ is dashed blue, $|C_{-1}(\tau)|^2$ is dotted red, $|C_{-2}(\tau)|^2$ is long dashed yellow, and $|C_{m}(0)|^2=\frac{1}{5}$ for all $m$ substates.  Top:  $\omega=\omega_0$.  Middle:  $\omega=\omega_0+\omega_1$. Bottom: $\omega=\omega_0+3\omega_1$.}}
  \end{figure}
       \begin{figure}
  \center{\includegraphics[width=0.49\textwidth]{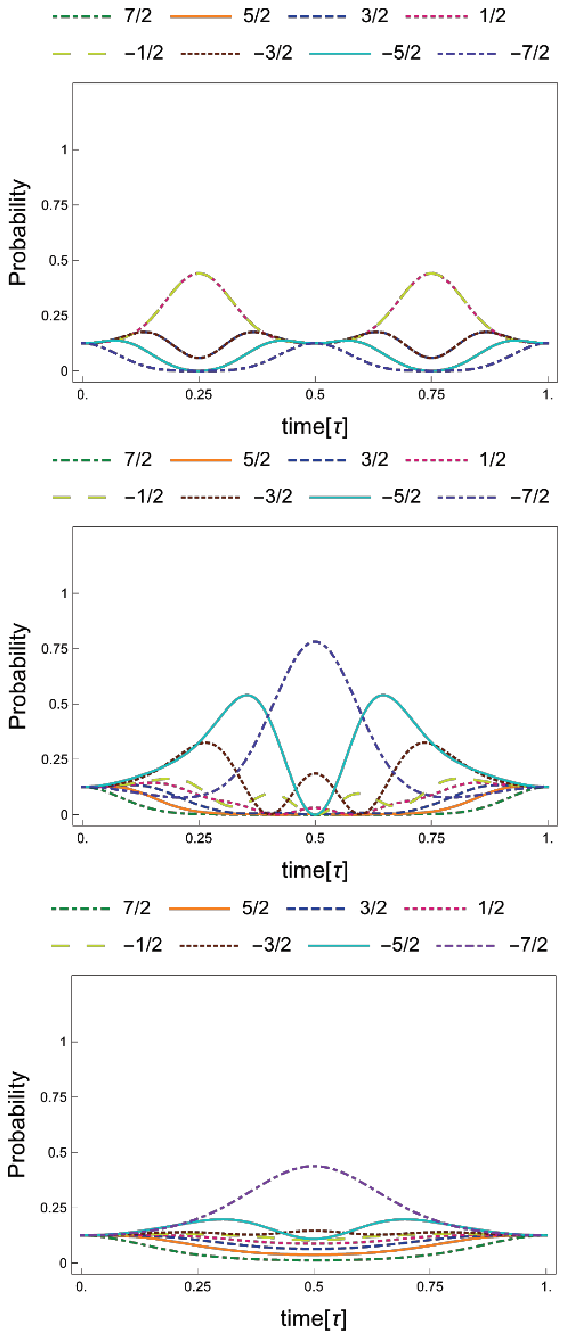}
  \caption{Plots for $s=\frac{7}{2}$ of the $|C_m(\tau)|^2$ from Eq. (\ref{probability}), where  the reduced time variable $\tau=t/T$, the period $T=2\pi/\Omega$, and $\Omega$ is given by Eq. (\ref{Omega}).  In each figure, $|C_{7/2}(\tau)|^2$ is dot dashed green, $|C_{5/2}(\tau)|^2$ is solid orange,  $|C_{3/2}(\tau)|^2$ is dashed blue, $|C_{1/2}(\tau)|^2$ is dotted red, $|C_{-1/2}(\tau)|^2$ is long dashed yellow, $|C_{-3/2}(\tau)|^2$ is short dashed brown,  $|C_{-5/2}(\tau)|^2$ is solid light blue,  $|C_{-7/2}(\tau)|^2$ is dot dashed purple,  and $|C_{m}(0)|^2=\frac{1}{8}$ for all $m$ substates.  Top:  $\omega=\omega_0$.  Middle:  $\omega=\omega_0+\omega_1$. Bottom: $\omega=\omega_0+3\omega_1$.}}
  \end{figure}
 \begin{figure}
  \center{\includegraphics[width=0.49\textwidth]{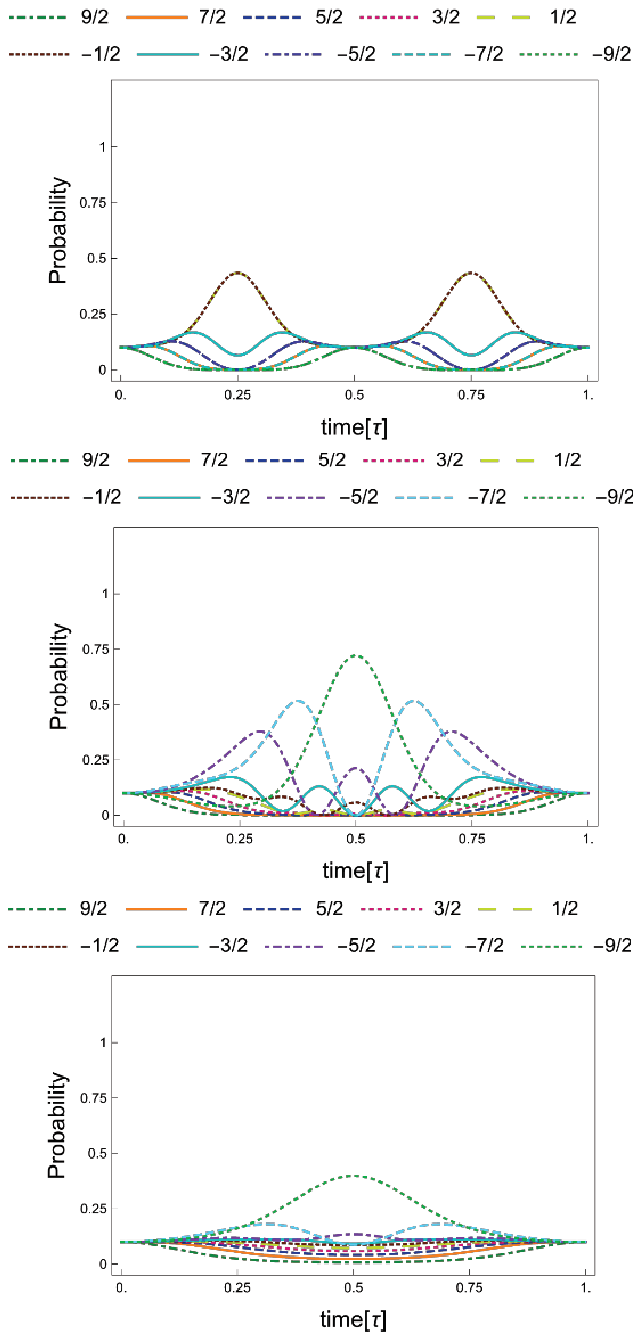}
  \caption{Plots for $s=\frac{9}{2}$ of the $|C_m(\tau)|^2$ from Eq. (\ref{probability}), where  the reduced time variable $\tau=t/T$, the period $T=2\pi/\Omega$, and $\Omega$ is given by Eq. (\ref{Omega}).  In each figure, $|C_{9/2}(\tau)|^2$ is dot dashed green, $|C_{7/2}(\tau)|^2$ is solid orange,  $|C_{5/2}(\tau)|^2$ is dashed blue, $|C_{3/2}(\tau)|^2$ is dotted red, $|C_{1/2}(\tau)|^2$ is long dashed yellow, $|C_{-1/2}(\tau)|^2$ is short dashed brown,  $|C_{-3/2}(\tau)|^2$ is solid light blue,  $|C_{-5/2}(\tau)|^2$ is dot dashed purple, $|C_{-7/2}(\tau)|^2$ is dashed light blue,  $|C_{-9/2}(\tau)|^2$ is dotted green, and $|C_{m}(0)|^2=\frac{1}{10}$ for all $m$ substates.  Top:  $\omega=\omega_0$.  Middle:  $\omega=\omega_0+\omega_1$. Bottom: $\omega=\omega_0+3\omega_1$.}}
  \end{figure}

The authors acknowledge helpful discussions with Talat S. Rahman, Luca Argenti, and James Harper.
 R. A. K. was partially supported by the U. S. Air Force Office of Scientific Research (AFOSR) LRIR \#18RQCOR100, and the AFRL/SFFP Summer Faculty Program provided by AFRL/RQ at WPAFB.


\begin{thebibliography}{99}
\bibitem{Stone} N. J. Stone, {\it Table of Nuclear Dipole and Electric Quadrupole Moments}, Atomic Data and Nuclear Data Tables {\bf 90}, 75-176 (2005).
\bibitem{Majorana} E. Majorana, {\it Atomi Orientati in Campo Magnetico Variable}, Nuovo Cimento {\bf 9}, 43-50 (1932).
\bibitem{Schwinger} J. Schwinger, {\it On Angular Momentum}, Nuclear Development Associates (1952).
\bibitem{RabiRamseySchwinger} I. I. Rabi, N. F. Ramsey, and J. Schwinger, {\it Use of Rotating Coordinates in Magnetic Resonance Problems}, Rev. Mod. Phys. {\bf 26}, .167-171 (1954).
\bibitem{Ramsey} N. R. Ramsey, {\it Molecular Beams}, (Clarendon Press, Oxford, UK 1956), App. E pp 427-430.
\bibitem{Gottfried} K. Gottfried, {\it Quantum Mechanics Vol. I: Fundamentals}, (W. A. Benjamin, Inc., New York, NY 1966).
\bibitem{Griffiths} D. J. Griffiths and D. J. Schroeter, {\it Introduction to Quantum Mechanics}, 3$^{rd}$ Ed., (Cambridge University Press, Cambridge, UK, 2018).
\bibitem{SN3} J. J. Sakurai and J. Napolitano, {\it Modern Quantum Mechanics}, (3$^{rd}$ Ed., Cambridge University Press, Cambridge, UK, 2021).
\bibitem{GY} K. Gottfried and T.-M. Yan, {\it Quantum Mechanics: Fundamentals}, (2$^{nd}$ Ed., Springer Science + Business Media New York, NY 2003).
\bibitem{HallKlemm} B. E. Hall and R. A. Klemm, J. Phys.: Condens. Matter {\bf 28}, 03LT01 (2016).
\end{thebibliography}
 \end{document}